\documentclass[onecolumn,epjc3]{svjour3}
\usepackage{mathtext,bbm,amsmath,amsfonts,amssymb,indentfirst,syntonly,graphicx,bm}
\smartqed  
\RequirePackage{graphicx}
\RequirePackage{subfigure}
\RequirePackage{xcolor}
\def\bc{\begin{center}}
\def\ec{\end{center}}
\def\be{\begin{eqnarray}}
\def\ee{\end{eqnarray}}

\journalname{Eur. Phys. J. C}

\begin{document}
\title{Reconciling the cosmic age problem in the $R_h=ct$ Universe}
\author{H. Yu \thanksref{addr1}
         \and
           F. Y. Wang\thanksref{e1,addr1,addr2}
}
\thankstext{e1}{Corresponding author: fayinwang@nju.edu.cn}
\institute{School of Astronomy and Space Science, Nanjing
University, Nanjing 210093, China \label{addr1} \and Key Laboratory
of Modern Astronomy and Astrophysics (Nanjing University), Ministry
of Education, Nanjing 210093, China \label{addr2} }

\date{Received: date / Accepted: date}
\maketitle

\begin{abstract}
Many dark energy models fail to pass the cosmic age test. In this
paper, we investigate the cosmic age problem associated with nine
extremely old Global Clusters (GCs) and the old quasar APM
$08279+5255$ in the $R_h=ct$ Universe. The age data of these oldest
GCs in M31 is acquired from the Beijing-Arizona-Taiwan-Connecticut
system with up-to-date theoretical synthesis models. They have not
been used to test the cosmic age problem in the $R_h=ct$ Universe in
previous literature. By evaluating the age of the $R_h=ct$ Universe
with the observational constraints from the type Ia supernovae and
Hubble parameter, we find that the $R_h=ct$ Universe can accommodate
five GCs and the quasar APM 08279+5255 at redshift $z=3.91$. But for
other models, such as $\Lambda$CDM, interacting dark energy model,
generalized Chaplygin gas model and holographic dark energy model,
can not accommodate all GCs and the quasar APM 08279+5255. It is
worthwhile to note that the age estimates of some GCs are
controversial. So, unlike other cosmological models, the $R_h=ct$
Universe can marginally solve the cosmic age problem, especially at
high redshift.

\end{abstract}

\section{Introduction}

Many astronomical observations, such as type Ia supernovae (SNe Ia)
\cite{Riess,Perlmutter,Schmidt}, the cosmic microwave background
(CMB) \cite{Spergel,Komatsu,Planck}, gamma-ray bursts
\cite{Wang11a,Wang11b} and large-scale structure (LSS)
\cite{Tegmark}, indicate that the Universe is undergoing an
accelerated expansion, which suggests that our universe may have an
extra component like dark energy. The nature of dark energy is still
unknown, but the simplest and most interesting candidate is the
cosmological constant \cite{Carroll}. This model can consist with
most of astronomical observations. The latest observation gives that
the present cosmic age is about $t_0=13.82$ Gyr in $\Lambda$CDM
model \cite{Planck}, but it still suffer from the cosmic age problem
\cite{Yang,Wang}. The cosmic age problem is that some objects are
older than the age of the universe at its redshift $z$. In previous
literatures, many cosmological models have been tested by the old
quasar APM $08279+5255$ with age $2.1\pm0.3$ Gyr at $z=3.91$
\cite{Friacas,Komossa}, such as the $\Lambda$CDM
\cite{Friacas,Alcaniz2003a}, $\Lambda(t)$ model \cite{Cunha}, the
interacting dark energy models \cite{Wang}, Generalized Chaplygin
gas model \cite{Alcaniz2003b,Wang2009}, holographic dark energy
model \cite{Wei}, braneworld models \cite{Movahed,Alam,Pires} and
conformal gravity model \cite{Yang2013}. But all of these models
have a serious age problem except the conformal gravity model, which
can accommodate this quasar at $3\sigma$ confidence level
\cite{Yang2013}.

In this paper, we will use the old quasar APM 08279+5255 at redshift
$z=3.91$ and the 9 extremely old Global Clusters
\cite{Ma2009,WangAj2010} to investigate the cosmic age problem in
the $R_h=ct$ Universe. The data of these 9 extremely old GCs listed
in Table 1 is acquired from the Beijing-Arizona-Taiwan-Connecticut
system with up-to-date theoretical synthesis models. The
evolutionary population synthesis modeling has become a powerful
tool for the age determination \cite{Tinsley68,Searle73}. In
\cite{Ma2009,WangAj2010}, they get the ages of those GCs by using
multi-color photometric CCD data and comparing them with up-to-date
theoretical synthesis models. But the ages of GCs derived by
different authors based on different measurements using same method
are not always consistent \cite{Ma2009}. We find that the 9 of those
GCs can give stronger constraints on the age of universe than the
old quasar APM $08279+5255$. Those 9 extremely old GCs have been
used to test the cosmic age problem in dark energy models in
previous work and many dark energy models have a serious age problem
\cite{Wang}. The $R_h=ct$ Universe is a cosmic model which is
closely restricted by the cosmological principle and Weyl's
postulate \cite{Melia2007}. In the $R_h=ct$ Universe, the
gravitational horizon $R_h$ is always equal to $ct$. The $R_h=ct$
Universe can fit the SNe Ia data well \cite{Melia2012} , explain the
growth of high-$z$ quasars \cite{Melia2013}, account for the
apparent absence in CMB angular correlation \cite{Melia2014}. As we
discuss above, many cosmological models can not pass the age test.
But whether the $R_h=ct$ Universe suffers the cosmic age problem in
still unknown.

The structure of this paper is as follows. In section 2, we
introduce the $R_h=ct$ Universe. In section 3, we give the
constraints on the $R_h=ct$ Universe from SNe Ia and H(z) data. Then
we will test the $R_h=ct$ Universe with the 9 extremely oldest GCs
and the old quasar APM $08279+5255$. The age test in other
cosmological models is presented in section 4. Conclusions will be
given in section 5.

\section{The $R_h = ct$ Universe}\label{sec:RhModel}
The $R_h = ct$ Universe is a cosmic model which is closely
restricted by the cosmological principle and Weyl's postulate
\cite{Melia2007,MeliaShevchuk}. For a certain age of universe $t$,
there is a limiting observable distance $R_h(t)$, which is called
cosmic horizon. Any signal beyond cosmic horizon can not be observed
by us. The horizon is defined as
\begin{equation}\label{R_h}
R_h = \frac{2GM(R_h)}{c^2},
\end{equation}
where $M(R_h)$ is the total mass enclosed within $R_h$
\cite{Melia2007,Melia2009}. From Eq. (\ref{R_h}), we can find that
cosmic horizon is a Schwarzschild radius. If we set the matter density
is $\rho$, then $M(R_h)=4{\pi}{R_h^3}{\rho}/3$, so it yields
\begin{equation}\label{R_h*}
R_h = \frac{3c^4}{8{\pi}G{\rho}}.
\end{equation}
The expansion of the universe is calculated from Friedmann equation
\begin{equation}\label{FriedmannEq}
H^{2}=(\frac{\dot{a}}{a})^2=\frac{8{\pi}G{\rho}}{3c^2}-\frac{kc^2}{a^2},
\end{equation}
where $H=\dot{a}/a$ is Hubble parameter, $a$ is scale factor, $k$ is
the spatial curvature constant and for $k=-1,0$ and $+1$ corresponds
an open, flat and closed universe, respectively. If we assume the
universe is flat, from Eq.(\ref{R_h*}) and Eq.({\ref{FriedmannEq}}),
we have $H=c/R_h$. For the $R_h = ct$ universe, we have $R_h = ct$.
We obtain
\begin{equation}\label{H*}
H=\frac{\dot{a}}{a}=\frac{1}{t},
\end{equation}
where $t$ is the age of universe. Solving Eq.(\ref{H*}) with
$a=\frac{1}{1+z}$ and initial condition $H=H_0$ when $z=0$, one can
get
\begin{equation}\label{H}
H=H_0(1+z).
\end{equation}
The luminosity distance in the $R_h=ct$ Universe is \cite{Melia2009}
\begin{equation}\label{d_L}
d_L=(1+z)R_h(t_0)\ln(1+z)=\frac{c(1+z)}{H_0}\ln(1+z),
\end{equation}
where $t_0$ is the age of the local universe. \

\section{Observational constraints on the $R_h=ct$ Universe}\label{sec:Constr}
In this section, we constrain the $R_h=ct$ Universe using the Union
2.1 SNe Ia data \cite{Suzuki} and the observed Hubble parameter data
$H(z)$. The SNe Ia distance moduli and the value of H(z) reported in
the literature are depend on the specific cosmological model, i.e.,
$\Lambda$CDM. When we use them to constrain other cosmological
models, the original data must be re-analyzed. Wei et al. (2014)
\cite{Wei2014} derived the SNe Ia distance moduli in the $R_h=ct$
Universe. For the Hubble parameter data, we choose 19
model-independent data from \cite{MeliaMaier2013}. Then we test the
model with the 9 extremely old GCs in M31 and the old quasar APM
$08279+5255$ based on the principle that all objects are younger
than its local universe.

\subsection{Constrain the $R_h=ct$ Universe with SNe Ia and $H(z)$ data}
SNe Ia are considered as the best standard candles to measure
distance and investigate the expansion of the universe. The Hubble
parameter $H(z)$ reveals the expansion of the universe directly. So
we use the SNe Ia and $H(z)$ data to constrain the $R_h=ct$
Universe.  The Union 2.1 sample contains 580 SNe Ia at redshift less
than 1.5 \cite{Suzuki,Kowalski,Amanullah}. Wei et al. (2014)
re-calculate those SNe Ia distance moduli and gives their redshift
$z_i$, distance modulus ${\mu}_{obs}(z_i)$ and its corresponding
error ${\sigma}_i$. The theoretical distance modulus is defined as
\begin{equation}\label{DisModThr}
{\mu}_{th}(z_i)=5\log_{10}d_L(z_i)+25.
\end{equation}
We can get theoretical distance modulus ${\mu}_{th}(z_i)$ for each
SN Ia from Eq.(\ref{d_L}). The ${\chi}^2$ for SNe Ia is
\begin{equation}\label{Chi2SNe}
{\chi}^2_{SN}(H_0) =
\sum\limits_{i=1}^{i=580}\frac{({\mu}_{th}(z_i)-{\mu}_{obs}(z_i))^2}{{\sigma}_i^2}.
\end{equation}
So ${\chi}^2_{SN}$ has only one parameter $H_0$. We can get the
best-fit $H_0$ by minimize ${\chi}^2_{SN}$ (see Table 2).
\cite{Melia2012} also found that the $R_h=ct$ Universe can well fit
the Union 2.1 sample.

The Hubble parameter values we use are obtained from previous
published literature
\cite{Jimenez,Simon,Gazta,Stern,Blake,Moresco,Zhang}. These Hubble
parameter data is complied in \cite{Farooq}. In
\cite{MeliaMaier2013}, 19 model-independent values have been chosen.
So we use these model-independent H(z) data. The $\chi^2$ for $H(z)$
is
\begin{equation}\label{Chi2H}
{\chi}^2_{H}(H_0) =
\sum\limits_{i=1}^{i=19}\frac{(H_{th}(z_i)-H_{obs}(z_i))^2}{{\sigma}_{H_i}^2}.
\end{equation}
The total ${\chi}^2$ is ${\chi}_{tot}^2(H_0) = {\chi}^2_{SN}(H_0) +
{\chi}^2_{H}(H_0)$. Then we minimize the total ${\chi}^2_{tot}$ to
get the best-fit parameter $H_0$ of the $R_h=ct$ Universe.

The best-fit Hubble constant is $H_0 = 70.01\pm0.40
{\rm~km~s^{-1}~Mpc^{-1}}$ at $1\sigma$ confidence level with
${\chi}_{min}^2=573.13$ from SNe Ia. After including the 19 Hubble
parameter data, the best-fit Hubble parameter is $H_0 =
69.83\pm0.40{\rm~ km~s^{-1}~Mpc^{-1}}$ at $1\sigma$ confidence level
with ${\chi}_{min}^2=604.03$. Recently, Planck team derives the
Hubble constant $H_0 = 67.3\pm1.2\rm~ km~s^{-1}~Mpc^{-1}$ in the
$\Lambda$CDM model, which is consistent with our result.

\subsection{Testing the $R_h=ct$ universe with old objects}

The old objects are usually used to test cosmological models,
especial the old high redshift objects \cite{Lima}. In previous
literatures, many cosmological models can not pass the cosmic age
test. We use the 9 extremely old GCs in M31 and the old quasar APM
$08279+5255$ to test the $R_h=ct$ Universe. Any object at any
redshift $z$ must be younger than the age of the universe at $z$,
i.e., $t_{obj}(z)<t_{cos}(z)$, where $t_{obj}(z)$ is the age of a
object at redshift $z$, and $t_{cos}(z)$ is the age of the universe
at redshift $z$. The age of a flat universe is given as
\cite{Alcaniz2003b}
\begin{equation}\label{tz}
t_{cos}(z)={\int}_{z}^{\infty}\frac{d\tilde{z}}{(1+\tilde{z})H(\tilde{z})}.
\end{equation}
From Eq.(\ref{H}), the age of the $R_h=ct$ Universe at redshift $z$
is
\begin{equation}\label{Rhtz}
t_{cos}(z)=\frac{1}{H_0(1+z)}.
\end{equation}

We use the best-fit value of Hubble constant $H_0 = 70.01\pm0.40
{\rm~km~s^{-1}~Mpc^{-1}}$ from SNe Ia data to calculate the age of
the universe. For this result, the age of local $R_h=ct$ universe
$t_0 = 13.97{\pm}0.08$ Gyr. For the best-fit value of Hubble
constant $H_0 = 69.83\pm0.40{\rm~ km~s^{-1}~Mpc^{-1}}$ from SNe Ia
and Hubble parameter data, the age of local universe is $t_0 =
14.01{\pm}0.08$ Gyr. We choose the second one. In Fig.
\ref{fig:Age_Ia_Hz_Q_fig}, the blue line shows the evolution of
cosmic age at different redshifts, and red lines are $1\sigma$
dispersion. For a given diagonal line, the area below this diagonal
line corresponds to a larger cosmic age. From Fig.
\ref{fig:Age_Ia_Hz_Q_fig}, we find that the $R_h=ct$ Universe
accommodates the old quasar APM $08279+5255$ at more than $3\sigma$
confidence level. In Fig. \ref{fig:Age_Ia_Hz_G_fig}, the blue line
shows the best fit line of the age of local universe, and red lines
are $1\sigma$ dispersion. From Fig. \ref{fig:Age_Ia_Hz_G_fig}, we
find that 5 GCs (B239, B144D, B260, B383, B495) can be accommodated
by the $R_h=ct$ Universe at $1\sigma$ confidence level but the other
4 GCs (B129, B024, B297D, B050) can not. But the age estimates of
some GCs are controversial. For example, the metallicities of B129,
B024, B297D and B050 measured by \cite{Ma2009,WangAj2010} are higher
than those of \cite{Galleti}. The values are significantly
different. So the GCs ages derived by \cite{Ma2009,WangAj2010} may
be larger than true ages. Due to the uncertainty of age
determination, we can claim that the $R_h=ct$ Universe can
marginally solve the cosmic age problem.

\
\section{Testing other models}\label{sec:OtherModels}
In order to compare with the $R_h=ct$ Universe, we also investigate
some other models. The theoretical luminosity distance is
\begin{equation}\label{lumDis}
d_{\rm L}=c(1+z)\int^z_0 \frac{d\tilde{z}}{H(\tilde{z})},
\end{equation}
where $H(z)$ is the Hubble parameter. Then we can use Eq.
(\ref{DisModThr}) to get the distance modulus. But the SNe Ia data
should be re-optimized for each model except the $\Lambda$CDM model,
which needs lots of work. So like previous literatures, we just use
the SNe Ia data based on the $\Lambda$CDM model. The 19
model-independent Hubble parameters chosen by \cite{MeliaMaier2013}
are also used.

\subsection{$\Lambda$CDM model}\label{sec:LCDM}
\begin{figure}
 \centering
 \subfigure[$\Lambda$CDM model.]{ \includegraphics[width=0.45\textwidth]{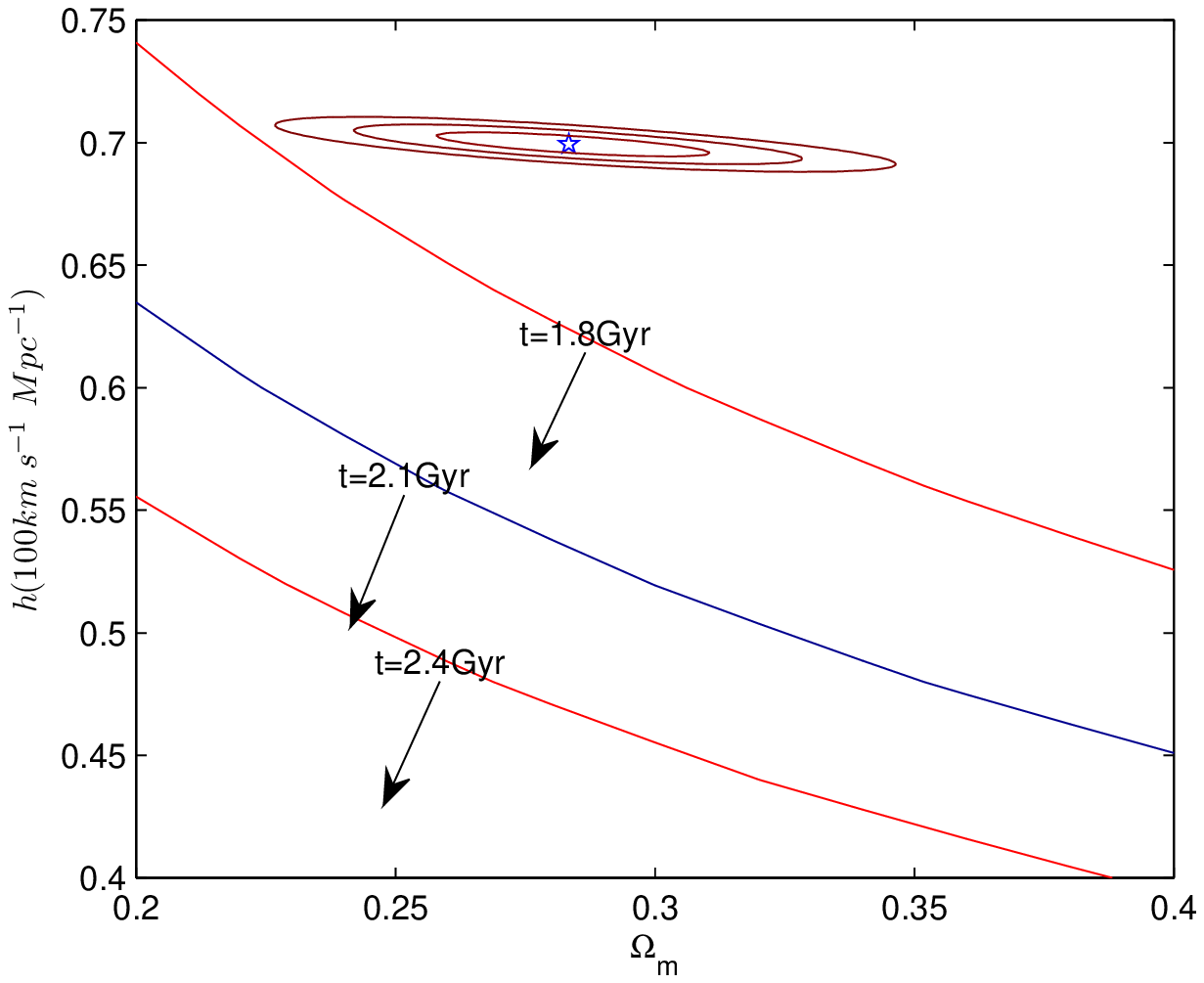}}
 \subfigure[I$\Lambda$CDM model.]{ \includegraphics[width=0.45\textwidth]{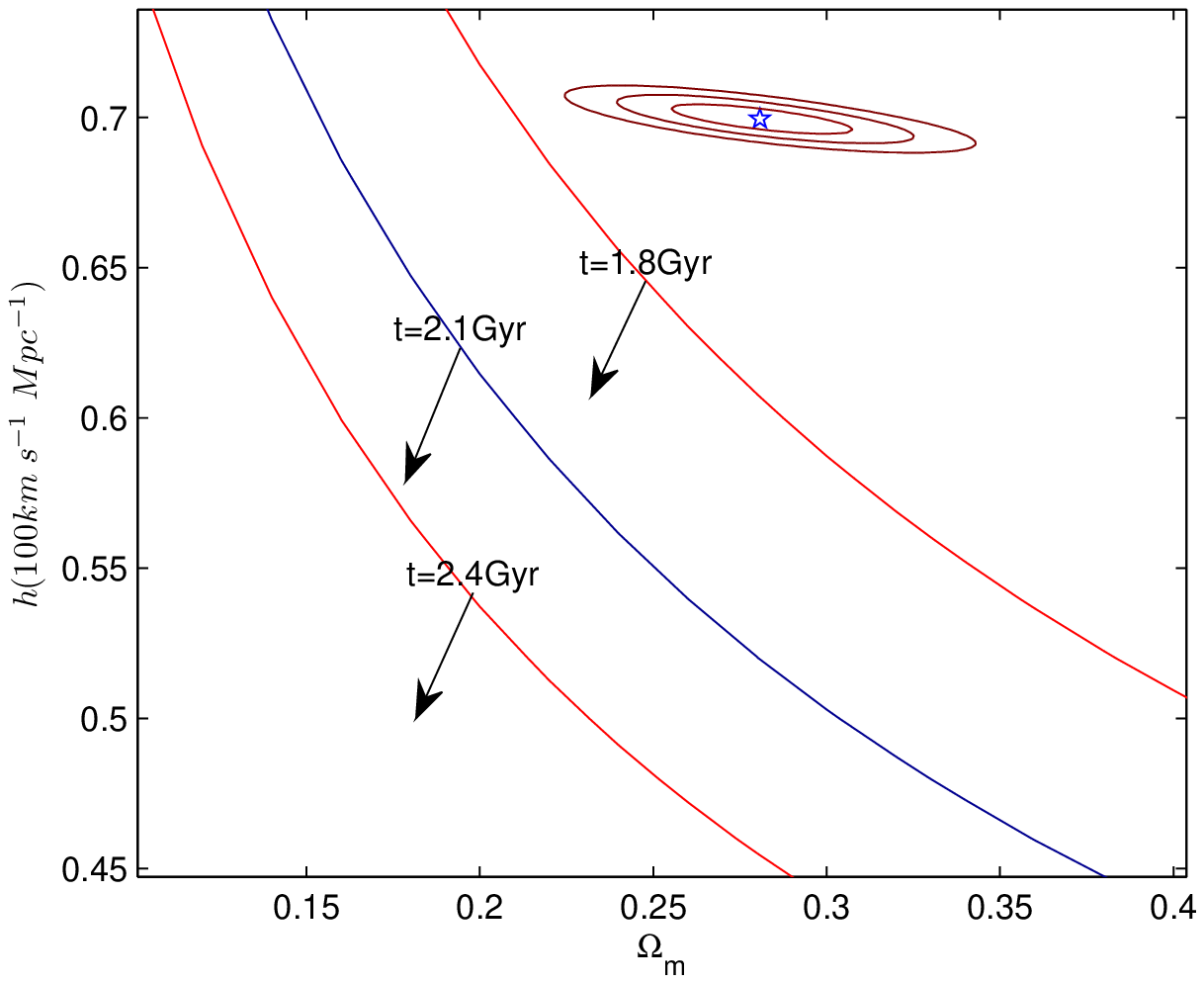}}
 \\
 \subfigure[The GCG model.]{ \includegraphics[width=0.45\textwidth]{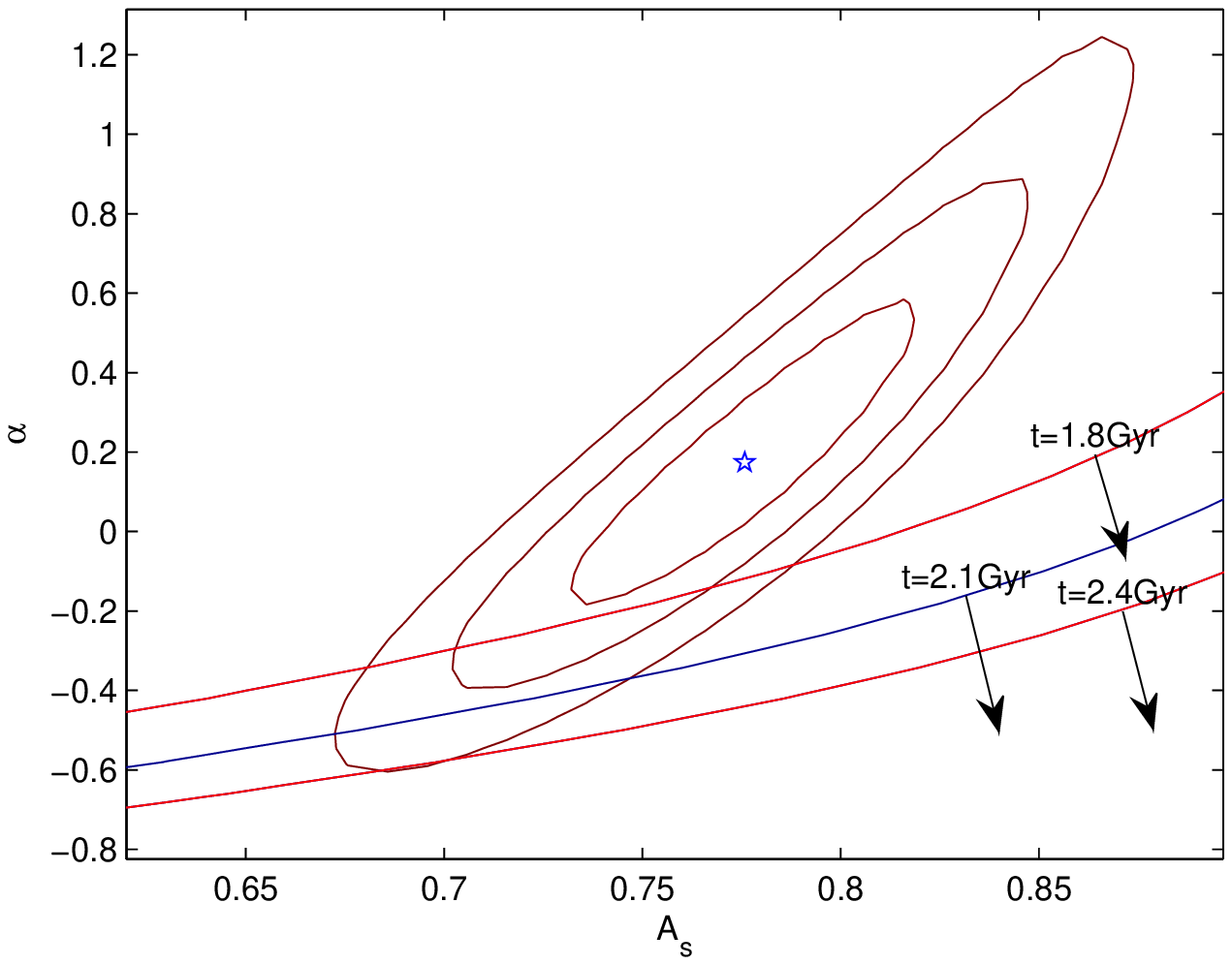}}
 \subfigure[The holographic dark energy model.]{ \includegraphics[width=0.45\textwidth]{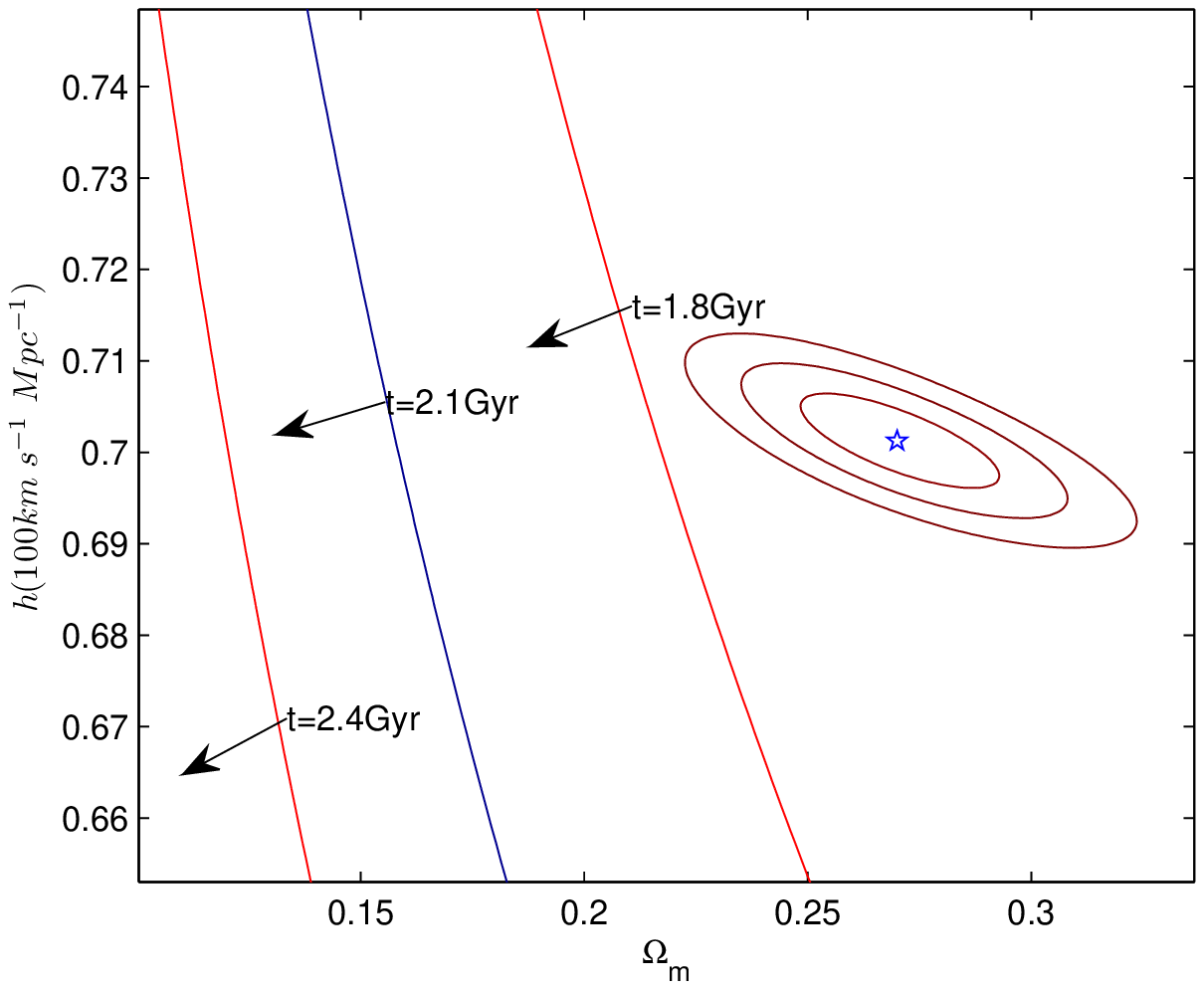}}

 \caption{Contour plot for $\Lambda \rm CDM$ model,  I$\Lambda$CDM model, the GCG model and the holographic dark energy model
 respectively. The ellipses represent confidence intervals
 from $1\sigma$ to $3\sigma$ and the the blue star means the optimal value. The blue line represents the age of universe
 at $z=3.91$ is 2.1 Gyr and the two red line represent the $1\sigma$ error $\pm 0.3$ Gyr. The arrowhead points the allowed region.}\label{contourFig}
\end{figure}

The Hubble parameter in the flat $\Lambda$CDM model is
\begin{equation}\label{Hz_LCDM}
H(z)=H_0\sqrt{\Omega_{m}(1+z)^3 + (1-\Omega_{m})}.
\end{equation}
Using the same method as that used in $R_h=ct$ model, we find the
best-fit Hubble constant value is $H_0 = 69.93\pm0.50
{\rm~km~s^{-1}~Mpc^{-1}}$ and the best-fit $\Omega_{m}$ value is
$\Omega_{\rm m}=0.28\pm0.02$. Panel (a) of Fig. \ref{contourFig}
shows the constraints on $h-\Omega_{m}$ plane at $1\sigma$,
$2\sigma$ and $3\sigma$ confidence level. The blue line and the two
red line represent the age of that old quasar APM $08279+5255$ and
$1\sigma$ error, respectively. From Fig. \ref{contourFig} we can see
that the $\Lambda$CDM model can not accommodate the old quasar APM
$08279+5255$. From Eq.(\ref{tz}), we can get the age of local
universe, which means $z=0$, is $t_0=13.71^{+0.30}_{-0.28}\rm~Gyr$.
From Fig. \ref{GCs_All}, which is similar with Fig.
\ref{fig:Age_Ia_Hz_G_fig}, we can also find that there are only 5
GCs (B239, B144D, B260, B383, B495) can be accommodated by the
$\Lambda$CDM universe at $1\sigma$ confidence level.

\subsection{Interacting dark energy model}\label{sec:ICDM}
In \cite{Wang}, they introduce three interacting dark energy models.
We take the first one called I$\Lambda$CDM as an example. For a flat
universe, the Hubble parameter in this model is
\begin{equation}\label{Hz_ILCDM}
H(z)=H_0\sqrt{\frac{\Omega_{
m}}{1-\alpha}(1+z)^{3(1-\alpha)}+(1-\frac{\Omega_m}{1-\alpha}}),
\end{equation}
where the $\alpha$ is a parameter denoting the strength of
interaction. The best-fit value are $H_0 = 69.95\pm0.50
{\rm~km~s^{-1}~Mpc^{-1}}$, $\Omega_{ m}=0.28\pm0.03$ and $\alpha =
-0.01$. From Panel (b) of Fig.\ref{contourFig}, we can see that the
I$\Lambda $CDM can not accommodate the old quasar APM $08279+5255$.
From Eq.(\ref{tz}), we can get the age of local universe is
$t_0=13.62^{+0.31}_{-0.27}\rm~Gyr$. From Fig. \ref{GCs_All}, we can
also find that there are only 4 GCs (B239, B144D, B260, B383) can be
accommodated by the I$\Lambda$CDM universe at $1\sigma$ confidence
level.

\subsection{Generalized Chaplygin gas model}\label{sec:GCGM}
\textbf{For the generalized Chaplygin gas (GCG) model, it has
\cite{Alcaniz2003b,Wang2009}:}
\begin{equation}\label{Hz_GCGM}
H(z)=H_0 \sqrt{\Omega_{b}(1+z)^3 +
(1-\Omega_{b})[A_s+(1-A_s)(1+z)^{3(1+\alpha)}]^{\frac{1}{1+\alpha}}},
\end{equation}
where $\Omega_{b}$ is the energy density of baryon matter, $A_s$ and
$\alpha$ are model parameters. The best-fit parameters are $H_0 =
70.07\pm0.35 {\rm~km~s^{-1}~Mpc^{-1}}$, $A_s=0.78\pm0.05$ and
$\alpha=0.17\pm0.38$. From panel (c) of Fig. \ref{contourFig}, we
can see that the age of the old quasar APM $08279+5255$ is in
tension (over $2\sigma$ confidence level) with the age of universe
for GCG model. The similar result is also found by \cite{Wang2009}.
From Eq.(\ref{tz}), we can get the age of local universe is
$t_0=13.73^{+0.38}_{-0.62}\rm~Gyr$. From Fig. \ref{GCs_All}, we can
also find that there are only 5 GCs (B239, B144D, B260, B383, B495)
can be accommodated by the GCG model at $1\sigma$ confidence level.

\subsection{Holographic dark energy model}\label{sec:HDM}
We will test the holographic dark energy model in this section. The
Hubble parameter in this model is \cite{Huang}
\begin{equation}\label{Hz_HDM}
H(z)=H_0\sqrt{\frac{\Omega_{ m_0}(1+z)^3}{1-\Omega_{ \Lambda}}},
\end{equation}
where $\Omega_{m_0}$ is the matter density at present and
$\Omega_{\Lambda}$ is the energy density of dark energy at redshift
$z$, which can be calculated by
\begin{equation}\label{OmegaL_HDM}
\ln{\Omega_{\Lambda}}-\frac{d}{2+d}\ln{(1-\sqrt{\Omega_{
\Lambda}})}+\frac{d}{2-d}\ln{(1+\sqrt{\Omega_{\Lambda}})},
-\frac{8}{4-d^2}\ln{(d+2\sqrt{\Omega_{\Lambda}})}=-\ln(1+z)+y_0,
\end{equation}
where $d$ is a free parameter and $y_0$ is a constant which can be
calculated by Eq.(\ref{OmegaL_HDM}) with $z=0$ and $\Omega_{
\Lambda}=1-\Omega_{\rm m_0}$. The best-fit parameters are $H_0 =
70.13\pm0.51 \rm~km~s^{-1}~Mpc^{-1}$, $\Omega_{ m_0}=0.27\pm0.02$
and $d=0.81\pm0.05$. From panel (d) of Fig. \ref{contourFig}, we can
see that the holographic dark energy model can not accommodate the
old quasar APM $08279+5255$. From Eq.(\ref{tz}), we can get the age
of local universe is $t_0=13.65^{+0.27}_{-0.26}\rm~Gyr$. From Fig.
\ref{GCs_All}, we can also find that there are only 4 GCs (B239,
B144D, B260, B383) can be accommodated by the holographic dark
energy model at $1\sigma$ confidence level.

\section{Conclusions}\label{sec:Conclusions}

\begin{figure}
\begin{center}
\includegraphics[width=\textwidth]{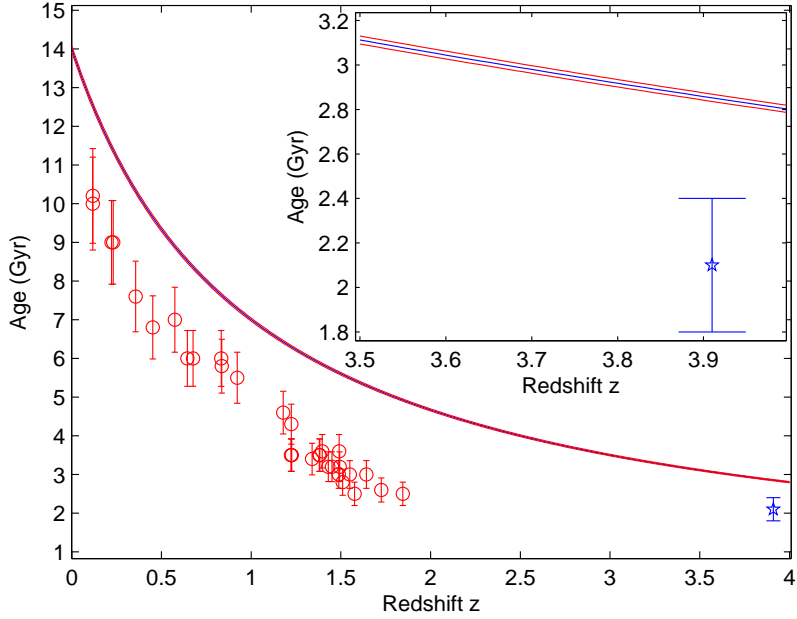}
\end{center}
\caption{The blue line shows the evolution of cosmic age in the
$R_h=ct$ Universe using the best-fit value from SNe Ia and Hubble
parameter, the red lines are the $1\sigma$ deviation. The star is
the old quasar APM 08279+5255. We can find the quasar are below the
lines, which means the old quasar APM 08279+5255 are younger than
the age of the $R_h=ct$ Universe. The open circle are old galaxies
data with $1\sigma$ error taken from \cite{Wei}. The insert shows
the the dispersion and data clearly. \label{fig:Age_Ia_Hz_Q_fig}}
\end{figure}

\begin{figure}
\begin{center}
\includegraphics[width=\textwidth]{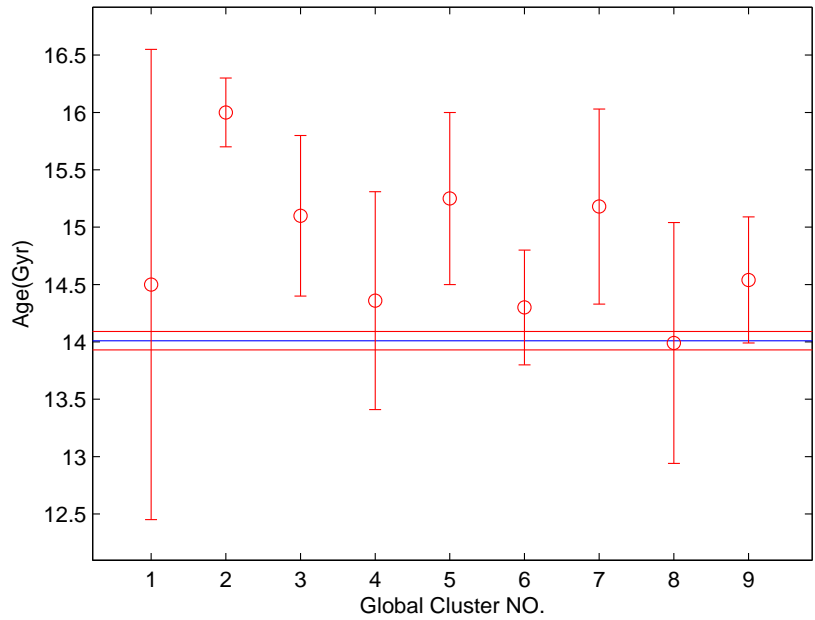}
\end{center}
\caption{The blue line shows the cosmic age in the $R_h=ct$ using
the best-fit value from SNe Ia and Hubble parameter, the red lines
are the $1\sigma$ deviation. The red circles are the 9 extremely old
GCs and it also gives the error of their
age.\label{fig:Age_Ia_Hz_G_fig}}
\end{figure}

\begin{figure}
 \centering
 \subfigure[$\Lambda$CDM model.]{ \includegraphics[width=0.45\textwidth]{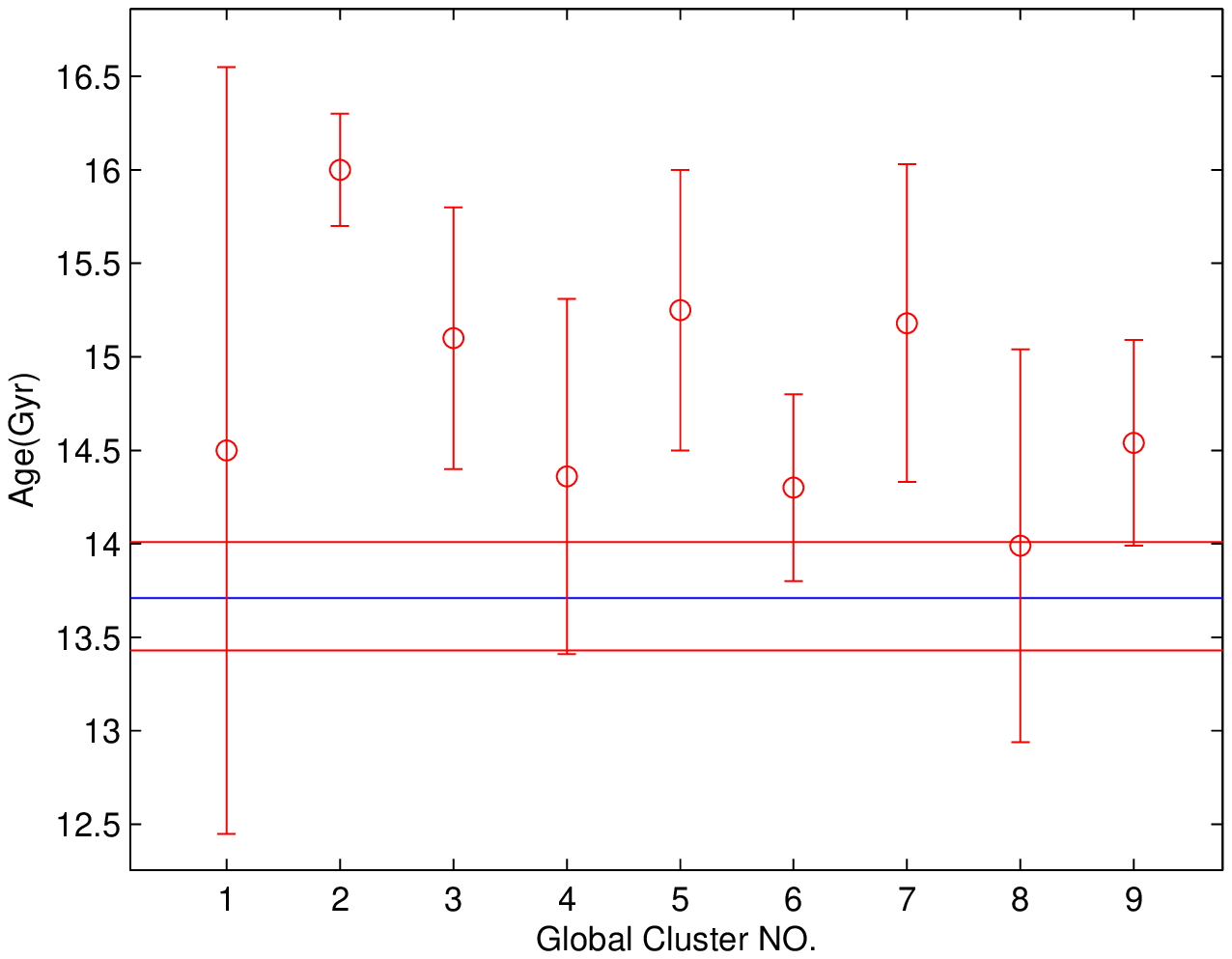}}
 \subfigure[I$\Lambda$CDM model.]{ \includegraphics[width=0.45\textwidth]{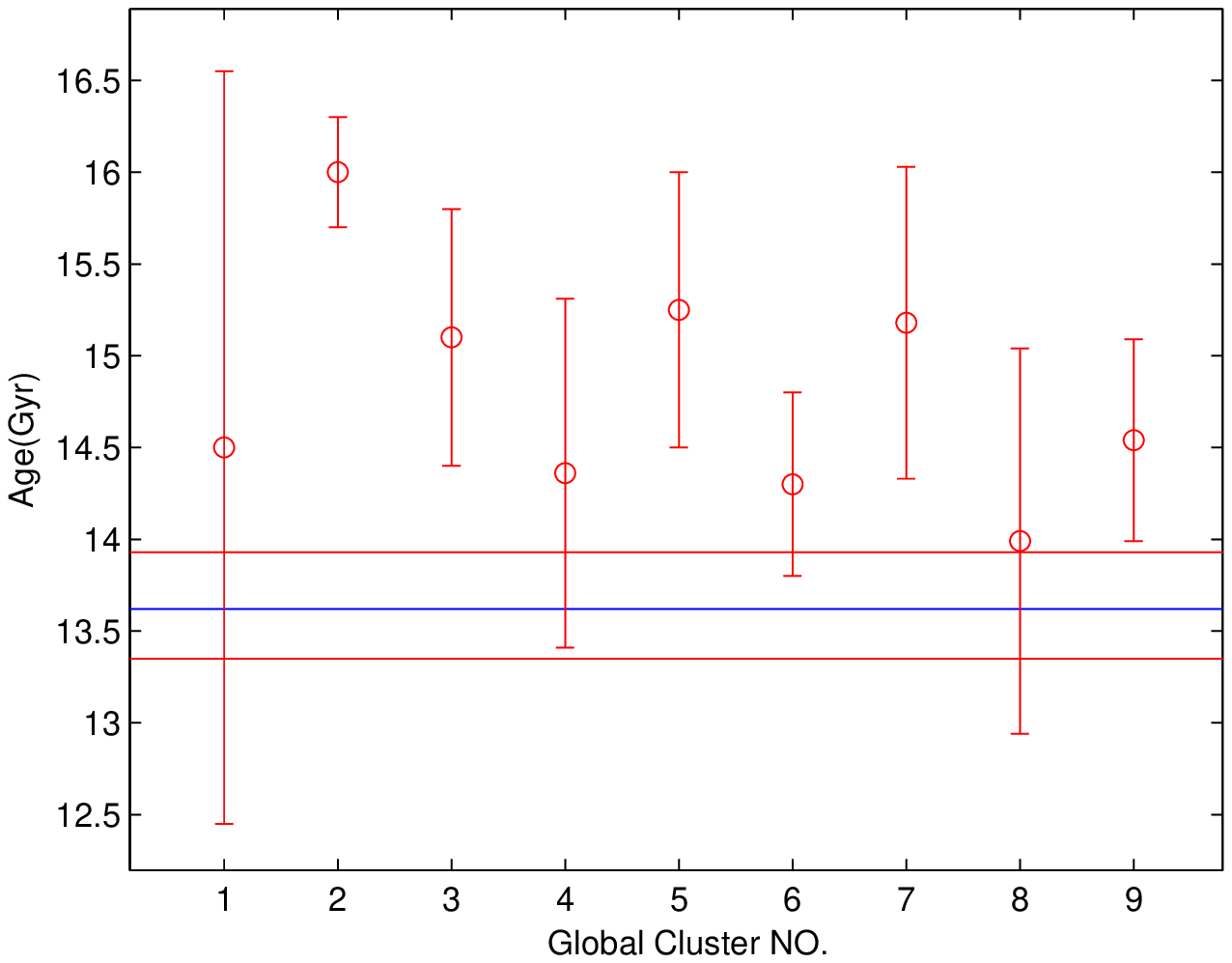}}
 \\
 \subfigure[The GCG model.]{ \includegraphics[width=0.45\textwidth]{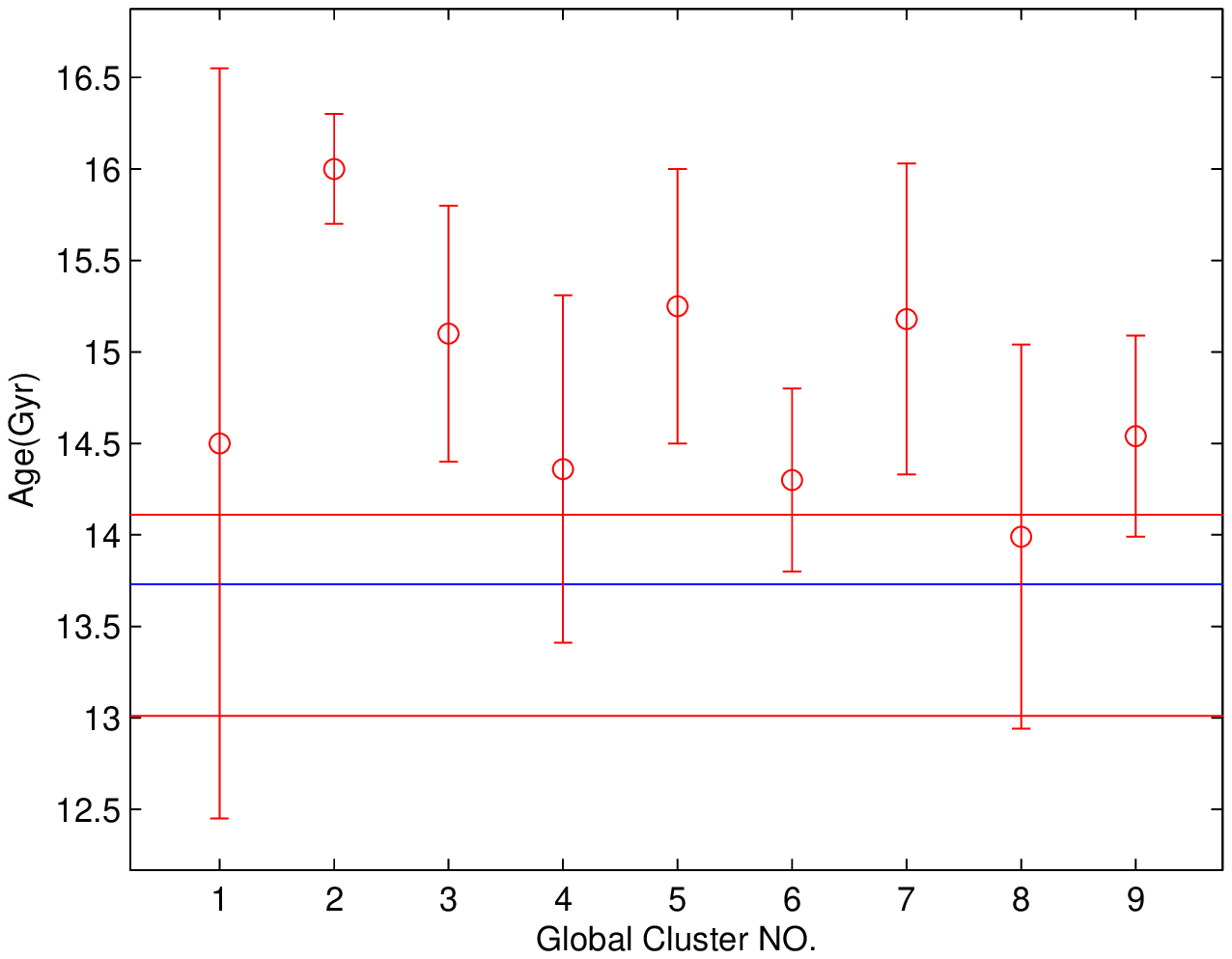}}
 \subfigure[The holographic dark energy model.]{ \includegraphics[width=0.45\textwidth]{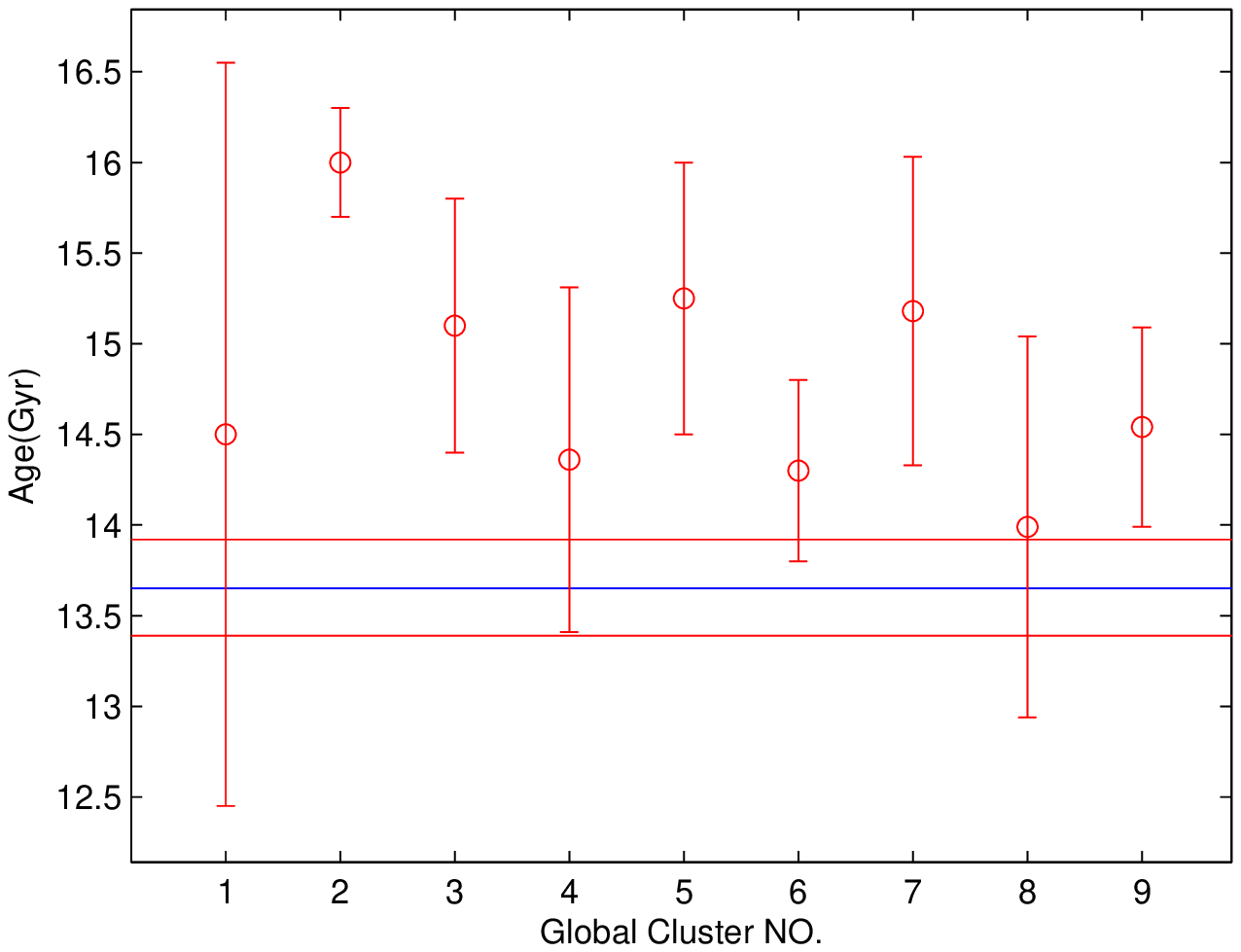}}
 \caption{Similar as Fig. \ref{fig:Age_Ia_Hz_G_fig} but for $\rm \Lambda CDM$ model, I$\Lambda$CDM model,
 the generalized Chaplygin gas model and the holographic dark energy model respectively.}\label{GCs_All}
\end{figure}

In this paper, we test the cosmic age problem in several
cosmological models by using nine extremely old GCs in M31 and the
old quasar APM $08279+5255$. We find that the best-fit value of
Hubble constant in the $R_h=ct$ Universe is $H_0 = 70.01\pm0.40
{\rm~km~s^{-1}~Mpc^{-1}}$ at $1\sigma$ confidence level by using SNe
Ia data. In this case, the age of local $R_h=ct$ universe $t_0 =
13.97{\pm}0.08 \rm Gyr$. If we fit the $R_h=ct$ Universe with the
SNe Ia and H(z) data, the Hubble constant is $H_0 =
69.83\pm0.40{\rm~ km~s^{-1}~Mpc^{-1}}$ at the $1\sigma$ confidence
level. The age of local universe is $t_0 = 14.01{\pm}0.08$ Gyr. From
Fig. \ref{fig:Age_Ia_Hz_Q_fig}, we find that the $R_h=ct$ Universe
can accommodate the old quasar APM $08279+5255$ at more than
3$\sigma$ confidence level. From Fig. \ref{fig:Age_Ia_Hz_G_fig}, we
find that there are five GCs (B239, B144D, B260, B383, B495) can be
accommodated by the $R_h=ct$ Universe at $1\sigma$ confidence level.
But the age estimates of some GCs are controversial. For example,
the metallicities of B129, B024, B297D and B050 measured by
\cite{Ma2009,WangAj2010} and \cite{Galleti} are significantly
different. So the derived ages are different. Due to the uncertainty
of age determination, we can claim that the $R_h=ct$ Universe can
marginally solve the cosmic age problem.

Using the same method, we also test some other cosmological models,
such as $\Lambda$CDM, interacting dark energy model, generalized
Chaplygin gas model and holographic dark energy model. In
Sec.(\ref{sec:OtherModels}), we show that these models can not
accommodate all nine old GCs in M31. Meanwhile, for the old quasar
APM $08279+5255$ at $z=3.91$, the $R_h=ct$ model can accommodate it
at more than $3\sigma$ confidence level. But these models can not
accommodate it. The generalized Chaplygin gas model is in tension
(over $2\sigma$ confidence level) with the age of APM $08279+5255$.
So the $R_h=ct$ Universe can marginally solve the cosmic age
problem, especially at high redshift.

\begin{acknowledgements}
We thank the anonymous referee for detailed and very constructive
suggestions that have allowed us to improve our manuscript. This
work is supported by the National Basic Research Program of China
(973 Program, grant No. 2014CB845800) and the National Natural
Science Foundation of China (grants 11373022, 11103007, 11033002 and
J1210039), the Excellent Youth Foundation of Jiangsu Province
(BK20140016), and the Program for New Century Excellent Talents in
University (grant No. NCET-13-0279).
\end{acknowledgements}


\newpage
\begin{table}
\centering
\begin{tabular}{c c c c}
\hline \hline
GC's NO. & GC & Age & Reference \\ \hline

1 & B239 & $14.50 \pm 2.05$ &  \cite{Ma2009} \\
2 & B050 & $16.00 \pm 0.30$ &  \cite{Ma2009} \\
3 & B129 & $15.10 \pm 0.70$ &  \cite{Ma2009} \\
4 & B144D & $14.36\pm 0.95$ &  \cite{WangAj2010} \\
5 & B024 & $15.25 \pm 0.75$ &  \cite{WangAj2010} \\
6 & B260 & $14.30 \pm 0.50$ &  \cite{WangAj2010} \\
7 & B297D & $15.18 \pm 0.85$ &  \cite{WangAj2010} \\
8 & B383 & $13.99 \pm 1.05$ &  \cite{WangAj2010} \\
9 & B495 & $14.54 \pm 0.55$ &  \cite{WangAj2010} \\ \hline

\hline
\end{tabular}
\caption{The properties of the 9 extremely old Global Clusters from \cite{Ma2009,WangAj2010}.
}
\end{table}

\begin{table}
  \centering
  \begin{tabular}{c c c }
    \hline \hline
    Observations & $H_0$/(km s$^{-1}$ Mpc$^{-1}$) & ${\chi}_{min}^2/dof$ \\ \hline
    SNe Ia & $70.01{\pm}0.40$ & 0.99 \\
    SNe Ia + H(z) & $69.83{\pm}0.40$ & 1.01 \\ \hline
    \hline
  \end{tabular}
  \caption{The best-fit values of the Hubble constant $H_0$ in the $R_h=ct$ Universe.}
\end{table}

\end{document}